\def\thin{{\thinspace}}

\def\ref{\par \noindent \hangindent=3pc \hangafter=1}

\def\sep{{\par \noindent \hangindent=15pt \hangafter=1}}

\def\scr{\scriptstyle}
\def\etal{{{\it et al}.\ }}

\def\pn{{\par\noindent}}

\def\rhobar{\lower 0.2ex\hbox{${\overline{\rho}}$}}

\def\;{;\thin}

\def\vectheta{{\rlap{$\theta$}{\hskip 0.1ex\hbox{$\theta$}}}}
\def\thetahat{{\rlap{\vectheta}{\hskip 0.2ex{\raise 0.5ex\hbox{\overline}}}}     }

\def\Pdot{{\hbox{$\dot P$}}}

\def\Mdot{{\hbox{$\dot M$}}}

\def\Mdot{{\hbox{$\dot M$}}}

\def\w{{\ \ \ \ \ }}
\def\tgs{{\thin \rlap{\raise 0.5ex\hbox{$\scr  {>}$}}{\lower 0.3ex\hbox{$\scr  {\sim}$}} \thin }}
\def\tls{{\thin \rlap{\raise 0.5ex\hbox{$\scr  {<}$}}{\lower 0.3ex\hbox{$\scr  {\sim}$}} \thin }}
\def\tll{{\raise 0.3ex\hbox{$\scr  {\thin \ll \thin }$}}}
\def\tgg{{\raise 0.3ex\hbox{$\scr  {\thin \gg \thin }$}}}
\def\tle{{\raise 0.3ex\hbox{$\scr  {\thin \le \thin }$}}}
\def\tge{{\raise 0.3ex\hbox{$\scr  {\thin \ge \thin }$}}}
\def\tl{{\raise 0.3ex\hbox{$\scr  {\thin < \thin }$}}}
\def\tg{{\raise 0.3ex\hbox{$\scr  {\thin > \thin }$}}}
\def\ts{{\raise 0.3ex\hbox{$\scr  {\thin \sim \thin }$}}}

\def\tp{{\raise 0.3ex\hbox{\fiverm +}}}
\def\Chi{{\raise 0.4ex\hbox{$\chi$}}}
\def\deg{{^\circ}}

\def\Msun{\hbox{$\thin M_{\odot}$}}

\def\Lsun{\hbox{$\thin L_{\odot}$}}

\def\Rsun{\hbox{$\thin R_{\odot}$}}

\def\a{$ ^a$}
\def\b{$ ^b$}

\def\ho{{H_{\rm orb}}}
\def\rht{R_{\rm HT}}
\def\tet{t_{\rm ET}}
\def\bp{B_{\rm P}}

\def\ttf{t_{\rm TF}}
\def\ra{R_{\rm A}}
\def\ro{R_{\rm o}}
\def\alf{Alfv{\'e}n}
\def\rl{R_{\rm L}}
\def\po{P_{\rm orb}}
\def\pr{P_{\rm rot}}
\documentclass[12pt,preprint]{aastex}
\newcommand{\be}{\begin{equation}}
\newcommand{\ee}{\end{equation}}
\begin{document}
\shorttitle{Cool Algols}
\title{The Evolution of Cool Algols}
\author{Peter P. Eggleton\altaffilmark{1,2} \& Ludmila Kiseleva-Eggleton\altaffilmark{1,3}}
\altaffiltext{1}{Institute of Geophysics and Planetary Physics, Lawrence Livermore National 
Laboratory, Livermore, CA 94550\\
    Email: {\tt ppe@igpp.ucllnl.org, lkisseleva@igpp.ucllnl.org}}
\altaffiltext{2}{On leave from the Institute of Astronomy, Madingley Rd,
         Cambridge CB3 0HA, UK}
\altaffiltext{3}{Dept. of Mathematics, St. Mary's College of California, Moraga, CA 94575}
\begin{abstract} 
Abstract: We apply a model of dynamo-driven mass loss, magnetic braking and
tidal friction to the evolution of stars with cool convective envelopes; in
particular we apply it to binary stars where the combination of magnetic
braking and tidal friction can cause angular-momentum loss from the {\it orbit}.
For the present we consider the simplification that only one component of a
binary is subject to these non-conservative effects, but we emphasise the need
in some circumstances to permit such effects in {\it both} components.
\par The model is applied to examples of (i) the Sun, (ii) BY Dra binaries, 
(iii) Am binaries,
(iv) RS CVn binaries, (v) Algols, (vi) post-Algols. A number
of problems regarding some of these systems appear to find a natural explanation 
in our model. There are indications from other systems that some coefficients
in our model may vary by a factor of 2 or so from system to system; this may
be a result of the chaotic nature of dynamo activity.
\end{abstract}

\section{Introduction}
In a previous paper (Nelson \& Eggleton 2001; Paper I) we constructed
a large grid of models of Case A binary evolution, according to the assumption
of conservative evolution. We found that these fitted reasonably the
parameters of certain observed `hot Algols', i.e. semidetached binaries in
which {\it both} components were earlier than about spectral type G0. We also
found that agreement was quite poor for some `cool Algols', by which
we mean those in
which at least one component is later than $\ts $G0. Since several of the
latter appeared to disagree on account of having less angular momentum,
and/or less total mass, than the theoretical models, we suggested that the
discrepancy is due to dynamo activity in stars with cool convective
envelopes. Such activity can reasonably be expected to be substantially
greater than in single stars of the same spectral type, because components
in Algols are typically rotating much faster than single stars, or stars
in binaries where the orbital period is quite long. This activity may reasonably
be expected to carry off both mass and angular momentum, but whether it can
carry off the right amounts is not clear, and is the main subject of this paper.
\par In another paper (Eggleton 2001; Paper II), PPE suggested a simplistic
model of dynamo activity, suitable for inclusion in a binary-evolution code.
In the present paper we present some results. The model is simplistic in the
sense that it gives the mass-loss rate, the overall poloidal magnetic field,
and the consequential magnetic braking rate, as functions of just four 
parameters: the mass, luminosity, radius and rotational period of the star. In
order that the magnetic braking should be able to drain angular momentum from
the {\it orbit}, it is necessary to include also a model of tidal friction. Of
course this is also necessary to get the star to spin faster in a binary than
it would if it were single. We follow the prescription of Eggleton \& Kiseleva
(2001: Paper III), but specialising to the simple case of only two bodies, and
stellar spin parallel to orbital spin.
\par In Section 2 we discuss briefly the implementation of these models in
a stellar evolution code, noting that there are some considerable
approximations which influence particularly low-mass systems, where quite
probably {\it both} components contribute to mass loss and angular-momentum
loss. In this paper we allow only the initially more massive star to be
subject to these processes. This is forced on us by numerical considerations,
but we hope to circumvent them in the future. In Section 3 we discuss our
results, and consider the further evolution of such systems.
In Section 4 we consider what may be required in future modeling.
\section{Incorporating non-conservative effects in a binary 
stellar-evolution code: the SNC model}
Several aspects of the evolution code used here have been described
elsewhere: we refer particularly to Paper I. Because of the non-Lagrangian
nature of the meshpoint distribution, it is particularly easy to add to
the mass {\it tranfer} rate (due to Roche-lobe overflow: RLOF) an additional
mass-loss rate due to stellar wind, provided it is a function only of surface
parameters such as mass, radius and luminosity, but we expect it to depend also
on stellar rotation rate $\Omega$, presumably through the Rossby number.
\par We have firstly to add to the usual
structure equations a further (rather trivial) differential equation for
the moment of inertia. We assume here {\it uniform} rotation in a star,
appealing to the argument of Spruit (1998) that internal magnetic field,
even if rather weak, is liable to wipe out such differential rotation as
might otherwise be expected when the core shrinks and the envelope grows. We
have to include $\Omega(t)$ as a variable to be solved for,
along with the variables normally determined by a stellar evolution code.
\par We further have to include two more unknown functions of time only that 
have to be solved for: the orbital angular momentum ($\ho$) and the 
eccentricity ($e$). The solution package in the code allows for the
possibility that along with the normal variables that change with position, 
i.e. pressure, temperature, luminosity, radius, mass and chemical composition 
in the case of a star, there may be a set of variables that are independent 
of position but which may nevertheless affect, and be affected 
by, the structure not just at the boundary but throughout the interior. 
The (uniform) rotation, for example, affects
the effective gravity throughout the star, using the standard von Zeipel
formulation in which the gravity is weakened by a factor $1-2\Omega^2r^3/3Gm$, 
where $r$ is the
volume-radius of an equipotential surface, and $m$ is the mass contained
within the equipotential. The quantities $\ho,e$ do not affect the interior
directly, in our approximation, but they are not known {\it a priori} since
they change in response to such physical processes as RLOF, tidal friction,
and ordinary nuclear evolution.
\par For the sake of a name, we call such quantities `eigenvalues', since like
eigenvalues they are constant in space but vary in time as the structure
varies. To our usual set of ten functions $m,r,L,p,T,X_1, X_4, X_{12},
X_{14}$ and $X_{16}$, i.e. mass, radius, luminosity, pressure, temperature
and 5 composition variables, and one eigenvalue (a constant which normalises
the meshpoint distribution), we therefore add one more function ($I$) and three
more eigenvalues ($\Omega, \ho$ and $e$). We have one more 
differential equation to add -- for $I(r)$ -- and four more boundary conditions.
One is $I=0$ at the centre, and the other three are equations at the surface
for the rates of change with time of $\ho,e$ and $\Omega$.
\par In the {\it conservative} approximation to RLOF it is very convenient
that it is possible to split the evolutionary calculation into two separate
parts. Firstly, we compute the evolution of star 1, allowing it to lose mass
if and when its radius exceeds its Roche-lobe radius. We can in principle
follow star 1 until it becomes either a white dwarf or a supernova, without
reference to the internal structure of star 2. The only information we need
about star 2 during this calculation is its mass, which is anyway uniquely
determined by the mass of star 1 and the assumption of conservative mass
exchange. Subsequently we can follow the evolution of star 2, giving it a
mass-gain rate which is the negative of star 1's loss rate. This separation
only becomes invalid when star 2 reaches its own Roche lobe -- as must
happen eventually. We then have to throw away the subsequent evolution of
star 1, which will no longer be valid; but at least we have a model for the
evolution of both stars up to this point.
\par This conveniently simple procedure unfortunately breaks down for
non-conservative evolution, if {\it both} stars are subject to stellar wind
and magnetic braking, since the orbit of star 1 will be varying at a rate 
partly due to the non-conservative processes on star 2. The best solution to 
this will be to follow both stars {\it simultaneously}, with each component
subject to boundary conditions that involve parameters at the surface of both
components. We are developing such a `Doubly-Non-Conservative' (DNC) code, 
but for the present we content
ourselves with a `Singly-Non-Conservative' (SNC) model, in which only star 1 
is subject to mass loss and magnetic braking. We can partly justify this on
the grounds that in many Algols star 2 (now the hotter and more massive
component) is of type A or B, and probably not as subject to dynamo activity
as star 1, which is typically a red giant or subgiant. We return to this point
subsequently.
\par Assuming, therefore, that only star 1 is subject to dynamo activity, we
can continue with the procedure outlined above: follow star 1 first, and 
subsequently star 2. The boundary conditions at the surface of star 1 are
written below, more-or-less directly in the form that they are computed in the
code. Since we are using an implicit technique, we are provided with initial
guesses of the following quantities at the surface: $M_1, R, L, I, \pr, \ho,
e$ and $M_2$. Note that we omit suffix 1 from all quantities relating to star 1
except mass, since, in regard to star 2, only the mass, and no other quantity, 
is involved. We then compute in order, acording to equations (1) -- (13) 
below, the following quantities:
\sep 1. $\rht$, a semi-empirical approximation to the Hayashi-track radius of
a star of mass $M$ and luminosity $L$: the ratio $R/\rht$ (0.55 for the Sun)
is a convenient measure of how convective a star is, and the fractional depth
of the surface convection zone is roughly proportional to its 2.1'th power
\sep 2. $\tet$, the envelope-turnover timescale for the convective envelope
\sep 3. $\ro$, the Rossby number, and $\Omega$, the intrinsic spin rate
\sep 4. $\zeta$, the mass-loss rate by dynamo-driven stellar wind
\sep 5. $\bp$, the mean poloidal magnetic field
\sep 6. $\ra$, the \alf\ radius of the stellar wind
\sep 7. The rate of loss of angular momentum from the {\it system}
\sep 8. $a$ and $\omega$, the semimajor axis and mean orbital frequency of
the binary; and $M$ and $q$, the total mass and the mass ratio
\sep 9. $\ttf$, an estimate of the timescale of tidal friction
\sep 10. The rate of change of eccentricity due to tidal friction
\sep 11. The rate of loss of angular momentum from the {\it orbit}: not the
same as quantity 7 above because angular momentum can be exchanged between the 
orbit and the stellar spin, through tidal friction, as well as carried 
away from the system by wind
\sep 12. $\rl$, the Roche-lobe radius of the star -- assuming, as is 
reasonable, that the orbit will have circularised before $\rl$ becomes
important
\sep 13. $\Mdot_1$, the mass-loss rate of star 1, a combination of RLOF if it
occurs and of stellar wind $\zeta$ above; and also $\Mdot$, the mass-loss rate
of the {\it binary}.
\pn In the following equations, $M$ is in units of $10^{33}\thin$gm, $R$ in 
units of $10^{11}\thin$cm, $L$ in units of $10^{33}\thin$erg/s, time 
(including inverse angular frequencies) in seconds, angular momentum in units
of $10^{55}\thin$cm$^2$gm/s and magnetic field in Gauss. 
These imply that the Newtonian constant $G$ is $10^{11}$ times it cgs value.
$$\rht={0.755L^{0.47}+0.05L^{0.8}\over M_1^{0.31}}\w,\eqno(1)$$
$$\tet=1.04\times 10^7\left({M_1R^2\over L}\right)^{1/3}
\left({R\over\rht}\right)^{2.7}\w,\eqno(2)$$
$$\ro={\pr\over\tet}\w,\w\Omega={2\pi\over\pr}\eqno(3)$$
$$\zeta=1.54\times 10^{-17}\thin{RL\over M_1}\thin\left({R\over\rht}\right)^2
(1+2.8\ro^2)^{-3.67}\w,\eqno(4)$$
$$\bp=54\left({M_1\over R^3}\right)^{1/2}\left({RL\over M_1}\right)^{1/3}
\left({R\over\rht}\right)^{3.4}(1+2.8\ro^2)^{-1.21}\w,\eqno(5)$$
$$\left({\ra\over R}\right)^{3/2}=2.5\times 10^{-36}\thin{R^5\over M_1}\thin
{\bp^4\over \zeta^2}\w,\eqno(6)$$
$${d\over dt}\thin(\ho+I\Omega)=-\zeta(\ra^2+{2\over 3}R^2)\thin\Omega-
\zeta{M_2\over MM_1}\ho\w,\eqno(7)$$
$$M\equiv M_1+M_2\ ,\w q\equiv{M_1\over M_2}\ ,\w {2\pi\over P}\equiv\omega=
{G^2M^5\over\ho^3}\thin{q^3\over(1+q)^6}\thin(1-e^2)^{3/2}\ ,\w 
a^3={GM\over \omega^2}\ ,\eqno(8)$$
$$\ttf=3\times 10^8\left({M_1R^2\over L}\right)^{1/3}
\left({a\over R}\right)^8{M_1^2\over M_2M}\w,\eqno(9)$$
$${de\over dt}= -{9e\over\ttf}\thin\left({1+{15\over4}e^2+{15\over8}e^4
+{5\over64}e^6\over\{1-e^2\}^{13/2}}-{11\Omega\over18\omega}\thin{1+{15\over4}e^2
+{1\over8}e^4\over\{1-e^2\}^5}\right) \w,\eqno(10)$$
$${d\ho\over dt}=-{\ho\over\ttf}\thin\left({1+{7\over2}e^2+{45\over8}e^4+{5\over16}e^6
\over\{1-e^2\}^{13/2}}-{\Omega\over\omega}\thin{1+3e^2
+{3\over8}e^4\over
\{1-e^2\}^{5}}\right)-\zeta{M_2\over MM_1}\ho\w,\eqno(11)$$
$$\rl=a\thin{0.49q^{2/3}\over0.6q^{2/3}+\ln(1+q^{1/3})}\w,\eqno(12)$$
$$\Mdot_1=-\zeta-1.58\times10^{-5}\left[\ln{R\over\rl}\right]^3\w,
\w\Mdot=-\zeta\w.\eqno(13)$$
In equation (13) the square brackets have a specific meaning:
$$[X]\equiv\max(0,X)\w,\eqno(14)$$
so that the RLOF mass-transfer rate is zero if $R\tle \rl$.
\def\st{St{\c e}pie{\' n}}
\par We emphasise the following points:
\sep (a) The constant factors in $\tet,\zeta,\bp$ and $\ra$ were chosen to give
roughly correct values for the Sun (Pizzo \etal 1983); the dependence on Rossby 
number of $\zeta$ and $\bp$ was chosen to give agreement with observational 
data on magnetic braking and stellar activity given by \st\ (1995) and 
Brandenburg, Saar \& Turpin (1998). The \alf\ radius is based on the theory of 
Mestel \& Spruit (1987). This is discussed in Paper~II.
\sep (b) Although many analyses of magnetic braking in binaries have used
Skumanich's (1972) relation between rotational period and age, we instead follow
\st\ (1995) in believing that Skumanich's relation gives too much braking both 
at short and at long periods. \st's and Skumanich's relations agree approximately 
in the period range $4 - 8\thin$d, but while Skumanich's relation implies a 
couple (a negative power of, in effect,
Rossby number) which diverges for $P,\ro\to 0$, \st's (a negative exponential)
tends to a constant for $P\tls 4\thin$d. However we feel that at {\it long} periods,
roughly longer than the Sun's ($26\thin$d), \st's exponential cuts off rather too strongly.
We have therefore adopted a Lawrentian form $(1+\alpha \ro^2)^{-\beta}$ which 
with suitable choice of $\alpha,\beta$ approximates \st's form for $\ro\tls 3$ while
decreasing less strongly at larger $\ro$. This Lawrentian form can be seen in equations
(4) and (5).
\sep (c) In equation (7), angular momentum loss is the sum of three terms. The
first is the {\alf}ic term, the second is due to the fact that even without
magnetic linkage the wind will carry off some spin angular momentum, and the 
third
is due to the fact that even if $\ra=0$ {\it and} $R=0$, i.e. the star is a point
mass, wind (assumed spherically symmetric, and fast) will carry off the same 
angular momentum per unit mass as resides in the orbital motion of star 1
\sep (d) The terms involving tidal friction (those which include $\ttf$)
come from the eqilibrium-tide theory (Hut 1981, Eggleton, Kiseleva \& Hut 1998). 
The timescale, as in $\tet$, has as 
a main factor the quantity $(M_1R^2/L)^{1/3}$, which has the dimensions of time
and can be seen as a turbulence-driven timescale whether or not the star is 
actually convective
\sep (e) If tidal friction is rather strong, equation (11) implies a transient
equilibrium between spin-up and spin-down. In the case that $\zeta=0$ this gives
the usual pseudo-synchronous period ratio $\Omega/\omega$ (Hut 1981), but the
ratio is modified if the mass-loss timescale ($M_1/\zeta$) is comparable to the
tidal-friction timescale
\sep (f) Equations (7), (10) and (11) are very `stiff' if the tidal-friction
timescale is short, as it is in {\it close} binaries. However this is no problem
numerically, because these equations, along with the stellar-structure equations,
are solved implicitly.
\sep (g) The mass-loss rate, emerging along with the magnetic field from an 
($\alpha,\omega$) model, turns out to resemble Reimers' (1975) empirical law,
although not by design. It differs in having a factor dependent on the
depth of convection, and another factor dependent on Rossby number. For
$R=\rht$, i.e. a fully convective star, and for $\ro\ts 3$, we obtain roughly
Reimers' law.
\pn Equations (7), (10) and (11) are the three extra surface boundary conditions
that are necessary to determine the three extra `eigenvalues' $\ho,\Omega$ and
$e$. Equation (13) is a boundary condition that was already included in the
conservative code (with $\zeta=0$, of course).
\section{Results}
We start by modeling the effects of mass loss, magnetic braking and tidal 
friction in a single star, and in {\it detached} binaries that have not yet evolved 
off the main sequence. We consider three problems:
\sep 1. The Sun
\sep 2. BY Dra and YY Gem, two low-mass close binaries where both magnetic braking 
and tidal friction are likely to play a role
\sep 3. Am binaries, where it has been suggested (Abt \& Bidelman 1969) that tidal friction
in binaries with $2.5\tls P\tls 100\thin$d may have slowed the spin of an A star enough
to allow the process of selective diffusion to create the observed composition
anomalies.
\subsection{The Sun}
If, following our prescription, we evolve a $1\Msun$ star which starts
with fairly rapid rotation (say $3\thin$d), we do not get very good agreement 
with the present-day Sun. This turns out to be entirely because our model
predicts a loss of mass of $\ts 1.4\%$ during the early rapidly-rotating
phase. We found good agreement (Table 1) if we started with $1.014\Msun$; at age 
$4.57\thin$Gyr the luminosity and radius agreed with measured values to better than
2\% and 0.1\% respectively. The rotational period was $24.8\thin$d, only about
5\% faster than observed. Agreement could no doubt be made exact by varying
fractionally the initial period (and consequentially the initial mass), the
metallicity ($Z=0.02$), and the mixing-length ratio ($\alpha=2.0$).
63\% of the mass loss occurred before age $0.25\thin$Gyr, by which time the 
rotational period had increased to $6.6\thin$d. The mass-loss rate at this 
point had dropped by less than a factor of 3 from its initial rate of 
$5\times 10^{-11}\Msun$/yr, but subsequently it dropped rapidly, as expected
from our Lawrentian form of dependence on Rossby number, reaching 
$3\times 10^{-14}\Msun$/yr currently. It is of course no coincidence that
we can get the period and mass-loss rate about right, since these (and the
current \alf\ radius) were used to normalise our formulae. Our predicted
early loss of mass does not seem to contradict any obvious feature of the
Sun and solar system, but it may have some relevance to problems of
light-element abundances in the Sun.
\begin{table}
\begin{center}
\centerline{Table 1. The evolution of the Sun}
\vskip 0.1truein
\begin{tabular}{lllllll}
\tableline
\tableline
age (Gyr) &mass    &log $R$ &log $L$   &$\pr$(d) &$\Mdot$($\Msun$/yr) &$\Delta M_{\rm SCZ}$ \\
\tableline 
0.0       &1.0140  &-0.0509 &-0.1497   &2.94     &-5.32$\times10^{-11}$ &0.032   \\
0.25      &1.0052  &-0.0456 &-0.1198   &6.61     &-1.94$\times10^{-11}$ &0.032  \\
4.57      &1.0005  &0.0002  &0.0076    &24.8     &-3.14$\times10^{-14}$ &0.022  \\
\tableline
\end{tabular}
\end{center}
\vskip 0.1truein
Masses, radii and luminosities in solar units; the last column is the mass of the
surface convection zone.
\end{table}
\subsection{BY Dra and YY Gem}
BY Dra (Vogt \& Fekel 1979) is a $6\thin$d binary of two K dwarfs, with a 
rather high eccentricity ($e=0.5$). At least one component is spotted, and
rotates with a period of $4\thin$d. This period is rather surprising, because it
is {\it not} the pseudo-synchronous period that one might expect ($\ts 2\thin$d).
It is expected that pseudo-synchronism is reached rather quickly, well before
the orbit becomes circularised. Of course the system is young, and perhaps it has
not yet reached even pseudo-synchronism. We find however that the combination
of our tidal-friction and magnetic-braking models can allow this system to have
reached a fairly steady transient equilibrium in which magnetic braking, slowing
the star down, is balanced by tidal friction, speeding it up.
\par Although there is the possibility that one or both components have not yet
reached the ZAMS, we restrict ourselves in this paper to models where both
components `start' on the ZAMS. BY Dra's orbit is well determined 
spectroscopically,
but there are no eclipses. Consequently although the mass ratio $q=1.2$ is known,
the masses and radii are indeterminate. We adopt masses of $0.71 + 0.58\Msun$,
which assume $\sin^3i=0.1$ and which seem reasonably appropriate for the late
K spectral type. We started the system with orbital period and eccentricity 
both slightly greater than the present values, and with a
rotational period (for star 1 only) of $2\thin$d. Note that throughout
this paper `star 1' is the component which was {\it initially} the
more massive. For some observed systems there may be scope for
argument, but in later discussion we make our preference clear with
this convention.
\begin{figure}
\plotone{f1.epsi}
%\centerline{\psfig{figure=fig1.ps,height=3.1in,bbllx=30pt,bblly=400pt,bburx=560pt,bbury=740pt,clip=}}
{ Fig 1 (a) --  Evolution of the eccentricity (crosses), orbital period 
(log, circles) and rotational period of star 1 (log, plusses) in BY Dra, as
functions of (log) age. For a considerable stretch of time ($30-500\thin$Myr) the rotational
period is about $5\thin$d, in transient equilibrium between spin-down due to
magnetic braking and spin-up due to pseudo-synchronisation. (b) -- The same,
but with the \alf\ radius reduced by a factor of 0.55. The rotational period
in transient equilibrium is reduced to about $3\thin$d. In (a) the angular-momentum 
loss is sufficiently rapid to lead the system into coalescence; in (b) mass
loss is more important than angular-momentum loss, and the period lengthens.
In both panels, the effect of mass loss and angular-momentum loss from
the lower-mass component is ignored.}
\end{figure}
\par Fig 1a shows how we expect the orbital period, rotational period and 
eccentricity to develop, as functions of $\log\thin($age), according to the 
prescriptions of Section 2. We see that star 1 spins down rather rapidly (in 
$\ts 30\thin$Myr) to a period of about $5\thin$d, and then much more slowly, in 
transient equilibrium between the magnetic-braking torque and the tidal-friction 
torque, as the period and eccentricity diminish from their original values to 
reach the present values at about $\ts 300\thin$Myr. Fig 1b is a similar plot 
but with $\ra$ from equation (6) reduced by a factor of
0.55, so that magnetic braking is diminished but not mass loss or tidal 
friction. The transient-equilibrium period is about $3\thin$d. Since the transient
equilibrium lasts for a long time, while the initial spin-down (or spin-up, if we
start with an initial rotational period which is substantially larger) is rather rapid,
we feel that it is likely that the observed system has indeed reached its transient
equilibrium value, and therefore that the tidal-friction timescale and the 
magnetic-braking timescale are rather comparable in this system, as predicted by our
model.
\par We do not feel, however, that we can confidently renormalise our model on the
basis of this one experiment. We have already mentioned some substantial uncertainies,
principally in the masses and in the evolutionary state. If one or both components
are pre-main-sequence stars, then the radii could be substantially larger and several
parameters might be quite different; although there would probably still exist a
transient equilibrium that could be at a similar rotational period. However, to
investigate such a possibility we would have to have a clearer idea of the origin of
this (and other) close binaries. The stars might not be coeval if the binary formed
from some capture or exchange mechanism. At present there is no clear understanding of
the origin of {\it close} binaries.
\par The two cases plotted in Fig 1 show very different behaviour in the long-term
future. In our canonical model (Fig 1a), magnetic braking is strong enough for the 
orbital period to decrease strongly after about $3\thin$Gyr. The system will probably
become a contact binary at $\ts 6\thin$Gyr, and then, with further magnetic braking, a merged single star.
With the slightly reduced magnetic braking (Fig 1b) this does not happen. Indeed the 
orbital period ultimately {\it increases}, because the mass of star 1, and so of the 
system as a whole, decreases significantly. By the end of the run shown in Fig 1b, 
$M_1$ was reduced to $0.44\Msun$. Of course, this emphasises the absurdity of not 
allowing star 2 to be correspondingly non-conservative. Provided that both stars stay near 
the ZAMS, it should be impossible for star 1 to become less massive than star 2. 
\par YY Gem is another low-mass binary of short period ($0.81\thin$d); it is a distant 
part of the well-known sextuple system $\alpha$ Gem. Because it eclipses, the masses 
($0.62+0.57\Msun$; Leung \& Schneider 1978) and radii are much better determined, and reasonably 
consistent with ZAMS models. We started the system with these masses, an eccentricity 
of 0.3, and orbital and rotational periods of $2.4\thin$d. This 
system reached the present orbital period after $1\thin$Gyr. Such an outcome
is not inconsistent with the fact that $\alpha$ Gem contains two early A main-sequence 
stars. Their masses are not known, but are likely to be in the range $2 - 2.5\Msun$,
which would give main-sequence lifetimes of $1.2 - 0.7\thin$Gyr.
\par YY Gem should be currently decreasing its period on a timescale of $\ts 0.5\thin$Gyr. 
Sowell \etal (2001) found the period of YY Gem to be approximately constant over
the last $75\thin$yr. One can estimate crudely from their Fig 4 that the timing of
the eclipse did not depart systematically by more than $\ts 5\thin$min from their 
revised ephemeris over this period. This gives a crude lower limit to the timescale 
$t_{\rm P}\equiv P/\Pdot$ of $\vert t_{\rm P}\vert\tgs 0.07\thin$Gyr, which is certainly consistent 
with our estimate. We might note that (a) star 1 lost about $0.02\Msun$ during its
evolution, so that it should have been started with a slightly greater mass, and
(b) it is likely that star 2 contributes about as much to angular-momentum loss
as star 1, so that the current spinup may be twice as great, and the system should
have started somewhat wider still.
\par Although the difference between Figs 1a and 1b is due to a difference
in the coefficients used, very much the same dichotomy between shrinkage (because
angular-momentum loss dominates) and expansion (because mass loss dominates)
is found using the {\it same} coefficients, but varying such initial quantities
as the masses and the period. For example, with masses $1.0+0.63\Msun$ the orbit
shrinks to contact (in $\tls 10\thin$Gyr) if the initial period is $\tls 3\thin$d, 
and expands if it is $\tgs 5\thin$d.
\subsection{Tidal Friction and Am stars}
The appearance of Am characteristics in A stars is well known to correlate
with rotational velocity: for $V\sin i\tgs 80\thin$km/s A-type spectra are
generally `normal', while for slower velocities they often have Am peculiarities. 
$V\sin i \ts 40\thin$km/s is reasonably typical for Am stars (Preston 1974).
Abt \& Bidelman (1969) found that Am stars are typically in binaries with
$2.5\tls P\tls 100\thin$d, and that normal A stars are in either shorter 
or longer
orbits, or else single. They suggested that this is due to tidal friction. In
short-period binaries the star is unable to spin slowly enough because of tidal
friction, while in long-period binaries its spin remains at its natal rate,
which on this hypothesis is typically faster than the critical value above.
\begin{table}
\begin{center}
\centerline{Table 2. The influence of tidal friction on a rapidly rotating A star in a binary}
\vskip 0.1truein
\begin{tabular}{lllllll}
\tableline
\tableline
Zero-age orbital period (d)              &10  &20   &40   &80   &160  &320 \\
\tableline 
Age (Myr) when  $V\sin i= 40\thin$km/s\w &3.95   &42.5   &361    &874    &978    &985    \\
\tableline
\end{tabular}
\end{center}
\end{table}
\par We have computed a number of binaries with masses $2.0 + 1.6\Msun$, a range
of period from 10 to $320\thin$d, an initial eccentricity of 0.5, and an initial 
rotational period of $0.76\thin$d. We were pleasantly surprised to find that tidal
friction, according to our recipe of Section 2, can have a significant effect
even at a period of $80\thin$d, although only if the evolution is followed all 
the way across the main sequence. In Table 2 we give the orbital period and the 
age at which the rotational velocity decreased to $40\thin$km/s. The end of the 
main sequence, defined somewhat arbitrarily as the point where central hydrogen 
had dropped to 0.02, was $1171\thin$Myr. We see that at an orbital period of 
$40\thin$d the star spends $\ts 70\%$ of its lifetime rotating more slowly than 
$40$km/s, whereas at an orbital period of $\tgs 160\thin$d it spends only 
$\ts 15\%$ of its life rotating this slowly. Only at the shortest orbital 
period considered ($10\thin$d) did the orbit become fully circularised and 
synchronised, at $6.9\thin$d, before the end of the main sequence, although at 
$20\thin$d the eccentricity had been reduced to 0.16 at this point.
\par By contrast, with an initial orbital period of $3\thin$d, which rapidly
reduced to $2.4\thin$d by circularisation, the rotational velocity of the star
dropped below $40\thin$km/s only between ages 5 and $386\thin$Myr. It therefore
spent $\ts 30\%$ of its life below this rotation velocity, whereas those with
$P\ts 10 - 40\thin$d spent $\tgs 70\%$ of their lives rotating this slowly.
\subsection{Provisional summary}
\par Our formulation of Section 2 appears to give some reasonably
satisfactory agreement with a number of features of detached binaries
of main-sequence stars. It gives a rather plausible explanation of
the rotational period of the spotted component in BY Dra, and can
account for the relatively slow rotation of Am stars in binaries of
period up to $\ts 80\thin$d. We are emboldened to see whether it can
give significant results in more evolved binaries, including 
mass-transferring (Algol) binaries,
where mass loss and angular momentum loss can be expected to play a much 
bigger role.
\section{More evolved binaries}
\par We now consider some binaries that have undergone substantial
evolution, and contain a red subgiant or giant.
\subsection{RS CVn binaries}
\par Red giants or subgiants in close but detached binaries (commonly
RS CVn stars) are known
to be unusually active compared to single red giants, or to those in wide 
binaries. The interpretation of this is that the giant is forced to rotate
faster in a binary than it would if it were single. Single stars should
spin down strongly as they evolve, partly because of evolutionary expansion and
partly because of magnetic braking. Tidal friction however opposes this
spin down.
\begin{table}
\begin{center}
\centerline{Table 3. Some RS CVn Systems}
\vskip 0.1truein
\begin{tabular}{lllllllllll}
\tableline
\tableline
   Name\w   & Spectra     &$M_1$    &$ M_2$   &$P$    &$e$   &$R_1$  &$R_2$  &log$T_1  $&log$T_2$  & Reference\\
            &             &         &         &       &      &       &       &          &          &\\
RS CVn \w   &G9IV + F4    &1.44     &1.41     &4.80   &      &4.0    &2.0    &3.707     &3.817     & Popper 1988a\\
Z Her \w    &K0IV + F5    &1.31     &1.61     &3.99   &      &2.7:   &1.9:   &3.697     &3.806     & \w     "  \\
RZ Eri \w   &K2III + F5m  &1.62     &1.68     &39.3   &.35   &7.0    &2.8    &3.625     &3.810     & Popper 1988b\\
RW UMa  \w  &K4IV-V + F8  &1.45     &1.5      &7.33   &      &3.8:   &2:     &3.630     &3.795     & Popper 1980\\  
\tableline
\end{tabular}
\end{center}
\end{table}
\par Table 3 gives some observational data for four RS CVn stars. The 
uncertainties in the data are discussed in the papers cited; but see 
below for a rediscussion of the radii in Z Her. Effective temperatures are
either directly from these papers, or else from $B-V$ colours given there
and combined with the $\log T\ vs\ B-V$ Table 2 of Popper (1980). It
can be seen that Z Her in particular shows a remarkable anomaly: the
cooler and more evolved component has {\it less} mass than the companion,
despite the fact that it does not fill its Roche lobe. This was interpreted
(Eggleton 1986) as evidence that the activity of the giant is so
substantially enhanced by rotation that it is losing mass on a {\it nuclear}
timescale, probably 2 or 3 orders of magnitude faster than one would
expect for a single star in the same evolutionary state.
\begin{table}
\begin{center}
\centerline{Table 4. Possible evolutionary history of Z Her}
\vskip 0.1truein
\begin{tabular}{llllllllllll}
\tableline
\tableline
              & Age  &$ M_1$ &$ M_2$ &$\po$&$\pr$&$ e$ &$R_1$&$R_2$&log$T_1$&log$T_2$&$\ho$  \\
              &        &         &         &       &       &       &       &       &          &          &\\
              &0       &1.80     &1.61     &3.90   &2.0    &0.3    &1.53   &1.48   &3.927     &3.884     &13.80    \\
circularised  &157     &1.80     &1.61     &3.39   &3.39   &0.01   &1.59   &1.50   &3.924     &3.885     &13.81    \\
ML starts     &1275    &1.75     &1.61     &3.46   &3.46   &0      &2.81   &1.99   &3.825     &3.847     &13.58    \\
present       &1645    &1.31     &1.61     &3.96   &3.96   &0      &3.47   &2.33   &3.709     &3.818     &11.15    \\
RLOF starts   &1664    &1.20     &1.61     &3.56   &3.56   &0      &4.90   &2.38   &3.687     &3.817     &10.00    \\
RLOF ends     &1698    &0.28     &2.21     &21.4   &21.4   &0      &9.77   &2.39   &3.640     &3.948     &6.12     \\   
              &1994    &0.27     &2.21     &22.9   &8.26   &0      &.030   &3.95   &4.380     &3.932     &6.07     \\
\tableline
\end{tabular}
\end{center}
\vskip 0.1truein
Age is in Myr, periods in days,  masses and radii
in solar units, and $\ho$ in arbitrary units.
\end{table}
\par No other RS CVn binary shows quite so marked a mass anomaly, although
RZ Eri and RW~UMa show a marginal anomaly. But several RS CVns, including the
prototype, have mass ratios remarkably close to unity, and this itself is a
little surprising since one component is usually substantially more
evolved than the other. It could be that such systems started with say
a 5--10\% mass difference, which has been whittled down to nearly zero:
2\% in the case of RS CVn. 
\par Unlike in the conservative case, it is difficult in the non-conservative
case to estimate initial configurations that will lead to currently observed 
parameters. After some experimentation, we started a Z Her model with parameters 
($1.8+1.61\Msun, \po=3.9\thin$d, $\pr=2.0\thin$d, $e=0.3$): row 1 of Table 4.
The orbit circularised fairly rapidly (row 2).  Star 1 lost little
mass ($\ts 0.05\Msun)$ until its radius had increased from $\ts 1.5\Rsun$ on
the ZAMS to about $2.8\Rsun$ (row 3). Thereafter mass loss accelerated strongly. When $M_1$ had
dropped to $1.31\Msun$ (row 4) the period had increased to $3.96\thin$d, roughly the 
observed value. 
\par An apparent problem with our model for Z Her is that the current radii of both components
(row 4 of Table 4) are too large, compared with the entries of Table 3, by about 25\%.
Popper (p.c. 1990) has suggested that, because the system is only partially eclipsing,
its inclination is rather uncertain and may have been overestimated. A slightly lower 
inclination allows the radii to be larger. Such a readjustment would also account for
the apparently low luminosity ($5\Lsun$) of the F5 component as tabulated by Popper 
 (1988a) -- as low as expected for a completely unevolved star of the same mass.
\par Table 4 continues the evolution to the onset of RLOF (row 5) at $1664\thin$Myr, its
cessation (row 6) at $1698\thin$Myr, and well into the post-RLOF phase when star 1 is a
hot subdwarf (row 7). Because of the fact that the mass ratio is reduced from an initial 1.12
to 0.75 prior to RLOF, the mass transfer is relatively well-behaved, and {\it not}
the rapid hydrodynamic transfer (Paczy{\'n}ski 1967) expected because the loser has
a deep convective envelope. In the whole of the evolution the total mass decreased by $27\%$ and
the orbital angular momentum by $56\%$. The angular momentum loss was split approximately evenly
between the detached pre-RLOF evolution and the semidetached evolution, while the
mass loss was somewhat greater in the immediate pre-RLOF evolution than in the
semidetached phase.
\par The much wider system RZ Eri (Table 3) can also be reasonably comfortably fitted 
with the same model. We started with parameters ($1.75+1.68\Msun, \po=49\thin$d, 
$\pr=2.0\thin$d, $e=0.5$). The results are shown in Table 5. The system parameters are 
very little altered until star 1 increases its radius from $1.5\Rsun$ to $5.6\Rsun$ 
(row 2), except that pseudo-synchronism is reached substantially sooner.
Circularisation is just beginning. Subsequently, the orbit circularises and star 1
loses mass on much the same faster timescale. The eccentricity, period and mass of 
star 1 all reach approximately their observed values at $1688\thin$Myr (row 3). Star 2 also has
about the right radius at this point, but star 1's radius is modestly too large: $9\Rsun$
instead of $7\Rsun$. In further evolution, the orbit circularises once star 1 grows to
$12\Rsun$ (row 4), and RLOF begins when star 1 reaches $\ts 21\Rsun$, by which time its 
mass is already reduced to $0.63\Msun$ (row 5). RLOF continued until star 1 was reduced 
to $0.34\Msun$ (row 6), after which star 1 began, as our run ended, to shrink towards 
the hot-subdwarf region (row 7).
\begin{figure}
\plotone{f2.epsi}
%\centerline{\psfig{figure=fig2.ps,height=3.1in,bbllx=35pt,bblly=520pt,bburx=565pt,bbury=780pt,clip=}}
{ Fig 2. Possible evolution of RZ Eri. (a) -- The theoretical Hertzsprung-Russell diagram. 
Star 1 is the thin line, star 2 the thick line. RLOF takes place only during the interval 
at the top right when $\log T \tls 3.6$. (b) -- Stellar radii (lower branch) and Roche lobe
radii (upper branch) for star 1
(thin lines) and star 2 (thick lines) as functions of mass. Both stars start near the right, 
at $\log R \ts 0.2$. Star 1 reaches its Roche lobe only when its mass is reduced to 
$\ts 0.63\Msun$ by the stellar wind postulated.  (c) -- Eccentricity (asterisks), rotational period (dots) and
orbital period (plusses) as functions of age (Gyr). Their variations are explained in the text.}
\end{figure}
\par Fig 2 illustrates some further aspects of the evolution of RZ Eri. In the HRD (Fig 2a)
we note that the mass loss from star 1 prior to RLOF causes the upward march on the
giant branch to be temporarily reversed; however the luminosity only drops by $\ts 25\%$
while the mass is almost halved. The upward march is restored, at $L\ts 100\Lsun$ star
1 fills its Roche lobe, and at about $\ts 260\Lsun$ star 1 detaches again, shrinking
down to the SDB region. Fig 2b is a superposition of four curves: the stellar and 
Roche-lobe
radii of both components, as functions of mass. The stars start towards the right,
below the middle. By the time star 1 reaches its Roche lobe it is much less massive than 
star 2, and as a result the RLOF is on a very slow, roughly nuclear, timescale, and {\it not}
on the very short (hydrodynamic) timescale that would normally be expected of a loser
with a deeply convective atmosphere.
\par Fig 2c shows the evolution of eccentricity (asterisks) as well as of rotational
period (dots) and orbital period (plusses). Until $\ts 1.6\thin$Gyr, $e$ and $\po$ were
essentially constant, while $\pr$ increased steadily fron $2\thin$d (an arbitrary starting value)
to the pseudo-synchronous value of $18\thin$d. The evolution of $\pr$ between 16 and 
$17\thin$Gyr is quite complex, because several timescales become comparable: nuclear
evolution, mass loss, angular momentum loss and tidal friction. There is a brief drop
in $\pr$ at the `hook' on the terminal main sequence, then a substantial increase, in
two steps, due to evolutionary expansion, and then a decrease as tidal friction reasserts
itself, the larger stellar radius countering the fact that the nuclear timescale has
shortened. But tidal friction increases so rapidly that it starts to cicularise the
orbit, and so as the star converges back to the pseudo-synchronous value the 
psudo-synchronous
value itself converges to the synchronous value, which is larger ($34\thin$d).
\par The star remains in synchronism until it starts to shrink rapidly away from its
Roche lobe. Although the radius decreases by a large factor, the moment of inertia
does not because the shrinkage is confined to the thin envelope surrounding the
degenerate core, having only $\ts 1\%$ of the mass. Consequently the star spins up
by only about a factor of four as it shrinks to the subdwarf region. We should
emphasise that we have assumed {\it uniform} rotation throughout star 1, at all times.
\begin{table}
\begin{center}
\centerline{Table 5. Possible evolutionary history of RZ Eri}
\vskip 0.1truein
\begin{tabular}{llllllllllll}
\tableline
\tableline
              & Age    &$ M_1$   &$ M_2$   &$\po$ &$\pr$  &$ e$   &$R_1$  &$R_2$   &log$T_1$&log$T_2$&$\ho$  \\
&&&&&&&&&&&\\
              &0       &1.75     &1.68     &49.0  &2.0    &0.5    &1.50   &1.49    &3.914   &3.897   &29.7     \\
ML starts     &1676    &1.74     &1.68     &47.7  &29.1   &0.49   &5.56   &2.85    &3.702   &3.801   &29.5     \\
present       &1688    &1.62     &1.68     &39.6  &23.4   &0.35   &8.79   &2.87    &3.682   &3.800   &28.1     \\
circularised\ &1697    &1.45     &1.68     &34.1  &34.4   &0.01   &12.1   &2.89    &3.663   &3.799   &26.0     \\
RLOF starts   &1726    &0.63     &1.68     &43.9  &43.9   &       &20.7   &2.93    &3.603   &3.797   &13.5     \\
RLOF ends     &1741    &0.34     &1.89     &129   &129    &       &34.3   &3.10    &3.601   &3.832   &12.0     \\
              &1743    &0.34     &1.89     &129   &22.6   &       &5.55   &3.10    &4.008   &3.831   &12.0     \\
\tableline
\end{tabular}
\end{center}
\vskip 0.1truein
Age is in Myr, periods in days, masses and radii
in solar units, and $\ho$ in arbitrary units.
\end{table}
\par RW UMa (Table 3) is also moderately well explained within the context of our
non-conservative model. Although the mass deficit of star 1 is by no means as 
well-determined as in Z Her, it is presumably substantially smaller, and this is 
accounted for in our model by the fact that the period is substantially larger
so that the mass-loss rate is significantly reduced. 
We start with masses $1.68+1.5\Msun$, i.e. much the same mass ratio as in Z Her, 
on the basis that this is necessary to obtain the considerable evolution required 
in star~2 that is manifested by its radius. Then we find that star 1's mass is reduced to
its present value when its radius is $\ts 5\Rsun$, a little larger than the
observed value but probably within observational uncertainty.
\par In fact RS CVn itself is a rather harder example to fit with our model than the
three other systems of Table 3. This system presumably started with much the same
mass ratio as Z Her and RW UMa, since the present components show much the same 
degree of evolution, and of differential evolution between the two components as 
manifested by their radii. Probable initial masses would therefore be $\ts 1.6+1.41\Msun$.
Since the period is slightly greater than in Z Her we might expect slightly less mass loss from
star 1, but actually we seem to have {\it much} less mass loss, by a factor of $\ts 3$.
We have not been able to obtain these parameters with the present model: our best attempts
give about twice as much mass loss as is required.
\par We do not try here to produce a detailed fit to the observed systems, say by a 
least-squares approach, for the following three reasons:
\sep (a) The observational uncertainties are quite substantial: see the papers cited
in Table 3. Only for Z Her is the sign of the mass difference clearly significant. 
Given that the uncertainties are substantial, it would probably not be difficult to 
find somewhat better fits to Z Her (Table 4) and RZ Eri (Table 5), but this would not 
be a very strong confirmation of the model.
\sep (b) The theoretical model of Section 2 is not put forward as definitive. It 
might be thought of as containing several coefficients and exponents whose values are unknown
{\it a priori} and which could be determined by fitting computed models to observed
systems; but it is rather unlikely that there is any unique formulation which will
correctly determine the mean behaviour of such a chaotic process as dynamo activity.
By varying enough coefficients from system to system we could reasonably expect to 
fit almost any observed system.
\sep (c) We should probably not assume that all these stars have the same metallicity
as the Sun, but nevertheless we have not varied $Z$ from a value of 0.02.
\pn We content ourselves by noting that the formulation of Section 2, without
varying any of the coefficients or exponents that it contains, seems to describe
reasonably well some observed RS CVn systems, and may have to be varied by factors
of about 2 to account for some more.
\subsection{Semidetached binaries}
The first four Algols listed in Table 6 show very clearly that they must
have lost some angular momentum and/or mass. Although some periods are not especially 
short, the very small mass ratios ($q\ts 0.075 - 0.16$) mean that the systems
have low angular momentum. Any Algol must have evolved through an equal-mass
configuration, and the orbital period then, assuming {\it conservative} evolution, 
is so small for S Cnc -- R CMa that even unevolved components would have 
overfilled their outer, let alone their inner, critical potentials. In addition,
R CMa has such a low {\it total} mass that it is difficult to see how it could
have evolved at all in a Hubble time. This would require that the initial mass
ratio was almost as extreme in the opposite sense as the current value; and in
that case the expectation would be that instead of steady RLOF we would have
the hydrodynamic mass transfer of Paczy{\'n}ski (1967), probably leading to a 
common-envelope phase with spiral-in (Paczy{\'n}ski 1976) and ultimately a merger.
\begin{table}
\begin{center}
\centerline{Table 6. Some Algols and possible post-Algol binaries}
\vskip 0.1truein
\begin{tabular}{llllllll}
\tableline
\tableline
Name          & Spectra        &$M_1$&$M_2$&$P$  &$R_1$&$R_2$& Reference\\
&&&&&&&\\
DN Ori        &G5III + A0        &0.34:  &2.8    &13.0   &6.7    &2.4    & Etzel \& Olson 1995 \\
S Cnc         &G8III + B9.5V     &0.23   &2.4    &9.49   &5      &2.2    & Olson \& Etzel 1993\\
AS Eri        &K0 + A3           &0.2    & 1.9   &2.66   &2.2    &1.8    & Popper 1980\\
R CMa         &G8IV + F1         &0.17   & 1.07  &1.14   &1.15   &1.5    & Sarma \etal 1996 \\
RT Lac        &G9IV + G5IV:      &0.63   &1.57   &5.07   &4.6    &4.3    & Popper 1980\\
RZ Cnc        &K4 + K1           &0.54   &3.2    &21.6   &12.2   &10.2   &  "\\
AR Mon        &K3 + K0           &0.8    &2.7    &21.2   &14.2   &10.8   &  " \\
DL Vir        &K0-2 + A3V        &1.1:   &2.2:   &1.32   &2.4:   &1.8:   & Sch{\"o}ffel 1977 \\
DL Vir\a      &(K + A) + G8III   &3.3:   &1.9:   &2260:  &       &       &  " \\
$\theta$ Tuc  &F: + A7IV         &0.063\b&0.7\b  &7.10   &       &       & De May \etal 1998 \\
V1379 Aql     &SDB + K0III-IV    &0.30   &2.27   &20.7   & .05   &9.0    & Jeffery \& Simon 1997 \\
FF Aqr        &SDOB + G8III      &0.35   &1.4    &9.21   &0.16   &7.2    & Vaccaro 2002\\
AY Cet        &WD + G5IIIe       &0.55:  &2.1:   &56.8   &.012   &6.8    & Simon \etal 1985\\
V651 Mon      &SDOB + A5V        &       &0.007\b&16.1   &       &       & M{\'e}ndez \& Niemela 1981 \\ 
0957-666      &WDA + WDA         &.32    &.37    &.061   &       &       & Moran \etal 1997 \\
AA Dor        &SDO + 4kK         &0.25:  &0.05:  &0.26   &016:   &0.09:  & W{\l}odarczyk 1984\\
\tableline
\end{tabular}
\end{center}
\vskip 0.1truein
\pn \a\ Wider orbit of triple system
\pn \b\ Mass function, or $m_i\sin^3 i$ 
\end{table}
\par Although our model for Z Her in Table 4 lost more than half its angular momentum,
it still ended up with more angular momentum than the four Algols we are considering.
We therefore explore some shorter initial periods. Furthermore, compared with the four 
RS CVn systems of Table 3, these Algols probably started with larger mass ratios. We 
infer this because in the Algols star 2 seems to be less advanced in its evolution, with 
radius nearer to the expected ZAMS radius for the current mass, than in the RS CVns. 
Hence we try $q_0\ts 1.3 - 1.5$, rather than $q_0\ts 1.05 - 1.12$. We can also infer a 
somewhat greater fractional loss of mass during the evolution, as compared with RS CVns, 
because our formulation implies greater mass loss at shorter period, other things being 
equal. A shot in the dark gives a system (Table 7) which reasonably resembles DN Ori at 
a late stage in its evolution. We start with a circular orbit and synchronised rotation, 
since these conditions will be established very quickly.
\begin{table}
\begin{center}
\centerline{Table 7. Possible evolutionary history of DN Ori}
\vskip 0.1truein
\begin{tabular}{llllllllll}
\tableline
\tableline
                    &Age    &$M_1$  &$M_2$  &$\po$  &$\ho$  &$R_1$  &$R_2$  &log$T_1$  &log$T_2$\\
&&&&&&&&&\\
                    &0      &2.24   &1.58   &1.62   &12.8   &1.69   &1.48   &4.002     &3.877     \\
RLOF starts         &725    &2.18   &1.58   &1.61   &12.5   &3.65   &1.72   &3.871     &3.895     \\
RLOF ends           &734    &1.49   &2.25   &1.70   &12.4   &3.20   &1.87   &3.765     &3.994     \\
RLOF starts         &1072   &0.932  &2.25   &1.54   &7.91   &2.52   &2.56   &3.729     &3.950     \\
present; RLOF ends  &1201   &0.263  &2.68   &13.0   &5.55   &6.85   &3.36   &3.749     &3.970     \\
                    &1212   &0.262  &2.68   &13.1   &5.55   &.084   &3.48   &4.525     &3.964     \\
\tableline
\end{tabular}
\end{center}
\vskip 0.1truein
\pn Age is in Myr, periods in days, masses and radii
in solar units, and $\ho$ in arbitrary units.
\end{table}
Rows 2 and 3 are the beginning and end of a first phase of RLOF. The system then
detaches slightly, but returns to RLOF at row 4. At row 5 the system is fairly
similar to the presently observed DN Ori. It is also very close to the end of
RLOF, and in row 6 star 1 has shrunk to an O-type subdwarf. We did not follow
the system further, but in row 6 star 2 is very close to the end of its
main-sequence life, and will shortly expand rapidly to bring about {\it reverse}
RLOF. Although our suggested mass for star 1 is substantially less than that
Table 6, it is within the error estimate of Etzel \& Olson (1995), viz. 
$0.34\pm 0.10\Msun$. An uncertainty of this magnitude is probably fairly typical
of low-mass companions in this kind of system. Etzel \& Olson obtained a 
{\it slightly detached} solution, which could easily be consistent with our 
finding that it is right at the end of RLOF.
\par S Cnc is rather similar to DN Ori: Olson \& Etzel (1993) found this system
to be also slightly detached.  Starting
values ($2.24+1.41\Msun, 1.51\thin$d) appear to be roughly adequate. At age
$1379\thin$Myr the corresponding parameters are ($0.25+2.53\Msun, 9.55\thin$d).
Star~2 is rather more evolved across the main-sequence band than we require.
Attempts to improve the starting parameters however were handicapped by the
fact there are at least five very different types of outcome that seem to
arise in quite a limited range of input parameter space. Three of these are fairly
similar to outcomes in conservative Case A, as described in paper I. There
we identified a total of 8 distinct outcomes, subCases of Case A which we
called AD, AR, \dots, AN. The three relevant here are Cases AR, AG and AL; but 
the fourth and fifth, with mass-loss and angular-momentum loss respectively being 
the dominant characteristic, have no analogy and we call them here Cases AM and AA.
\sep (i) AR -- Rapid evolution to contact: thermal-timescale RLOF, with a fairly 
large initial mass ratio, causes star 2 to expand rapidly to contact after rather 
little mass exchange.
\sep (ii) AG -- Giant contact: star 2 just misses the rapid contact of AR, but 
its substantial growth in mass allows it to catch up and overtake star 1's evolution 
so that contact between two red giants is reached at a later stage. 
\sep (iii) AL -- Late overtaking: star 2 misses rapid contact by a somewhat wider 
margin, but at a late stage, when star 1 is already a hot subdwarf, expands to fill 
its Roche lobe, initiating {\it reverse} mass transfer. This will no doubt be
on a dynamical timescale, leading to common-envelope evolution, spiral-in, and
either a merger (of the hot subdwarf with the white-dwarf core of star 2) or
to a close detached pair of white dwarfs. 
\sep (iv) AM -- Mass-loss dominated: star 1 loses mass sufficiently copiously that 
it never reaches its Roche lobe, although it may still evolve to a red subgiant and 
then a hot subdwarf. Alternatively there might be two minor episodes of RLOF 
separated by a substantial detached interval where wind alone causes the binary
to modify its period on about the same (nuclear) timescale as star 1
evolves. The evolution of DN Ori in Table 7 is of this character. Later
evolution will be the same as in Case AL.
\sep (v) AA -- Angular-momentum-loss dominated: star 1, and consequently the binary, 
loses angular momentum sufficiently rapidly that its Roche lobe shrinks rapidly to 
the stellar radius. Star 1 evolves down or very close to the ZAMS; usually star 2 
expands rapidly too, and the result is a contact system much as in the conservative 
Cases AR or AD (Paper 1).
\pn In Paper I we investigated the three-dimensional space of initial parameters
for {\it conservative} evolution with
several thousand models, to identify regions where the different Cases AD -- AN
occur. With our non-conservative model here the detailed behaviour even
in the much more limited domain that probably covers most of the systems in
Tables 2 and 5 seems more complex. We therefore content ourselves with
very qualitative agreement. It is possible that both DN Ori and S Cnc will
require that some coefficients in our non-conservative model be
modified, for example to give more angular momentum loss relative to mass
loss, but the case for this does not seem compelling.
\par AS Eri and R CMa are rather more difficult to account for with our
model. AS Eri has substantially less angular momentum than DN Ori or S Cnc;
but if we {\it start} with a substantially shorter period then the stars
are so close together that after RLOF
begins star 2 tends to expand quickly to a contact configuration (Case AR). Even if
this is just avoided, our formulation produces more mass loss at shorter 
period, and as a result the \alf\ radius is reduced and there is {\it less}
angular momentum loss, relative to the mass loss. We appear to need a formulation
with perhaps twice as much magnetic braking relative to mass loss as in our 
canonical model.
\par Some attempted models of AS Eri and R CMa had sufficient mass loss that 
RLOF was avoided altogether (Case AM), even though star 1 might be stripped 
down to a red subgiant and then a hot subdwarf core at a period not unlike 
the period of AS Eri. This raises a
potentially awkward question: can we be sure that all of these binaries
are indeed semidetached? It is normally taken as an {\it assumption} that a
system is semidetached, if the larger, cooler star is much less massive
than the smaller, hotter star. But if it is accepted that mass loss by
wind, from a star that does not yet fill its Roche lobe, can be on a nuclear 
timescale (as it must be at least in the case of Z Her) then the period and 
separation can increase on a nuclear timescale and so allow the windy star 
to remain a little smaller than its Roche lobe for a substantial period of time.
There may in practice be rather little difference in appearance between a
system where star 2 is accreting part of a wind from star 1, and one where
star 2 is accreting from RLOF of star 1 while star 1 is also losing mass
by wind. However, the measured parameters of such a system, particularly the
inclination, will depend on whether star 1's radius is 90\% or 100\% of its
Roche lobe radius.
\par We note that our model did not include the possibility of partial
accretion by star 2 of the wind from star 1. It would not in principle be
very difficult to include it, provided we had a formulation of the process
that we believed in. There is no doubt that such a process can take place:
for example, many though not all Ba stars must have accreted from the Ba-rich
stellar wind of an AGB-star companion, rather than from RLOF (Han \etal 1995).
But this process may be more effective for the copious, slow, cool winds of AGB
stars than the more meagre, fast and hot winds of red subgiants. A preliminary
estimate is that perhaps 10\% or less of the wind would be accreted, which we
consider small enough to be ignored.
\par The next three entries in Table 6, RT Lac, RZ Cnc and AR Mon, are all
double-(sub)giant binaries, examples of Case AG. The fact that star 2 is almost
as evolved as star 1 suggests that they started with more nearly equal masses 
(say $q_0\tls 1.05$, and perhaps $q_0\ts 1.01 - 1.02$)
than even the RS CVn systems, let alone the first four Algols. This presents 
us with the extra problem that they will almost certainly need a 
DNC model, and be less agreeable to our SNC model. 
We have therefore not attempted to model them. However, mass loss on a nuclear
timescale from star 1 can enlarge the region of Case AG, since it slows down
the evolution of star 1 and so makes it easier for star 2 to catch up.
\par DL Vir (Table 6) is a particularly interesting system because not only
is it triple, but the distant third body (star 3, say) is already evolved to 
the giant branch. This gives us the important information that star 1 of the 
close Algol pair must have had much the same initial mass as star 3. This in turn
poses an upper limit to the amount of mass lost by the system, since star 2
must have had less mass than star 1 initially. But at the same time there is
a lower limit, because if the initial mass ratio were close to unity star 2
would be more evolved than it is, as in the RS CVn systems ($q_0\ts 1.05 - 1.12$) 
and the three double-giant Algols. Taking the numbers in Table 6 at face value, 
we have in fact very little scope for mass loss: the starting masses must have 
been $\ts 1.9+1.6\Msun$, and the mass lost $\ts 0.2\Msun$. However the 
uncertainies in the observed masses are
considerable, and so there is probably also scope for as much mass loss
(relatively) as in our tentative model of DN Ori (Table 7), particularly as
the modest mass ratio in DL Vir puts it at a much earlier stage of mass
transfer.
\par Although several Algols are in triple systems (Chambliss 1992), DL Vir 
is the only one where star 3 is evolved to the giant branch, and thus affords
us a real estimate of the initial mass of star 1, not just a lower limit. 
We would hope that this important system could be re-analysed with modern 
technology.
\subsection{Post-Algols}
Table 6 contains two binaries, $\theta$ Tuc and V1379 Aql, that can be
recognised as probable post-Algols. Although the inclination of $\theta$ Tuc
is not known, a guess of $\sin^3i\ts 0.3 - 0.4$ makes it rather similar to
DN Ori and S Cnc, but presumably at a more advanced state of evolution where
star 1 has detached from its Roche lobe. The fact that we find it slightly
difficult to get periods as short as in S Cnc and DN Ori, and very difficult
to get periods as short as in AS Eri, may mean that $\theta$~Tuc requires
slightly enhanced angular momentum loss relative to our canonical model.
\par However V1379 Aql seems to fit fairly comfortably into the future 
evolution of Z Her, as indicated in Table 4. The only discrepancy is that
the $M_1$ that we end up with is about $10\%$ less than observed. That is
rather significant in this unusually accurately determined system. It will
not be easy to get rid of this by tinkering with initial parameters, since
the mass of the remnant is largely dictated by the size of its {\it current}
Roche lobe, i.e. by the size of the immediate red-giant precursor. The radius 
of a red giant is very sensitive to the mass of its core, and to get a core
10\% more massive should require a lobe almost twice as large. However
Jeffery (p.c. 1997) has suggested that the system may be slightly metal
poor, and this would certainly act in the right direction.
\par It is possible that the system is in fact losing angular momentum
currently, thanks to the activity of star 2 which has strong activity as in
RS CVns. As we have emphasised several times, we ignore the activity of
star 2. But it is unlikely to have reduced the orbital angular momentum
by as much as the mass discrepancy requires.
\par A very different anomaly in V1379 Aql is the fact that its orbit
is very significantly non-circular: $e=0.09\pm0.01$. A possible explanation,
though not our preferred one, is as follows. When RLOF ended, star 2 would
probably have been rotating rapidly, since the accretion stream from star 1
acquires angular momentum, because of Coriolis force, as it travels to star 
2. Equation (10) shows that if $\Omega\tg 18\omega/11$ (for $e\ts 0$) then
eccentricity {\it increases}. Tidal friction would probably have been
unimportant at first, but as star 2 grew to its present radius tidal friction
may have recently become important. This would require that even after some
evolutionary expansion, and also some angular momentum loss by magnetic
braking, star 2 was still rotating substantially faster than
$\ts 1.65\omega$. However the eccentricity would have to build up from
some small (but non-zero) value. We estimate that the Algol might have
finished RLOF with $e\ts 5\times10^{-5}$, on the following grounds.
Certain radio-pulsar plus WD binaries with comparable orbital period
show such an eccentricity, which is probably due (Phinney 1992) to 
small inhomogeneities of density and hence gravity on the
scale of turbulent convective elements in the red-giant loser before it
contracted to a WD. This would have to be amplified by a factor of
$\ts 2\times 10^3$ to reach the present value. It seems unlikely that
star 2 would have had enough angular velocity to do this, since the same
process that increases $e$ also decreases $\Omega$ to its corotational
value.
\par A more probable cause of the eccentricity, we believe, is the
action of a third body in a wide orbit highly inclined to the known
orbit. Such a body can drive long-period cycles, of both eccentricity
and inclination, in the $21\thin$d orbit (Kozai cycles; Kozai 1962). 
The amplitude of the eccentricity fluctuation does not depend on either 
the mass of the third body (which might therefore be a very inconspicuous 
M dwarf) or on its orbital period (which might be several years), but 
only on the inclination of the outer to the inner orbit. Such third bodies 
have recently been found for two systems (SS Lac, Torres \& Stefanik 1999; 
V907 Sco, Lacy \etal 1999), with longer orbits of $679\thin$d and $99\thin$d
respectively. These third bodies were needed to account for the fact that 
the close pairs sometimes eclipse, and sometimes do not, as a result of 
the fluctuation of inclination to the line-of-sight. Another Algol system, 
$\delta$ Lib, has also recently been found to have a third body (Worek 2001),
in a $1008\thin$d orbit. 
\par Whether such third-body orbits are typically highly inclined or not 
is at present unclear; but those of SS Lac and V907 Sco must be, in order
to cause the variation of inclination of the eclipsing orbit to the
line-of-sight that is observed. The prototype Algol has a third body
in an orbit inclined at $100\deg$ to the eclipsing orbit (Lestrade \etal
1993). Eggleton \& Kiseleva (2001) give equations governing the interaction 
of tidal friction, tidal distortion and Kozai cycles in SS Lac and other 
triples. They found that the inclination of the outer orbit to the inner in 
SS Lac had to $29\deg$. If hierarchical triples are typically caused by 
binary-binary interactions in a dense star-forming cluster, then inclinations 
higher than $60\deg$ are as likely as lower inclinations. An inclination of 
$60\deg$ would cause the eccentricity to oscillate between zero and 0.764. 
The period of the oscillation is $\ts P_{\rm out}^2/P_{\rm in}$, multiplied 
by a factor of total mass over $M_3$. Although the inclination and 
eccentricity fluctuate in the course of a Kozai cycle, the period does not.
\par The next three systems of Table 6, FF Aqr, AY Cet and V651 Mon are
potential post-Algols, because each has a hot-subdwarf star 1, a
less-evolved star 2, and an orbital period in the expected range.
However, although the masses of the hot components are all uncertain,
they appear to be on the high side: one at least appears to be more appropriate
to an AGB star than to an early FGB star. This suggests that they are products
of common-envelope evolution. Although the mass of the hot subdwarf in 
V651 Mon is unknown the system is the central star of the planetary nebula 
NGC 2346, which suggests a recent common-envelope origin. But the systems 
are somewhat surprising in that context
also since the usual expectation is that the common-envelope phase will
trigger spiral-in to a short period, a day or less, thus allowing the
system to evolve subsequently to a cataclysmic variable. In a tentative
study of common envelope evolution, to be published, we conclude that the
only common-envelope precursors to suffer {\it substantial} orbital shrinkage
are those with a mass ratio more severe than 4:1, in the sense of
(AGB star):(WD or MS companion). Systems with a less severe mass ratio
shrink their orbits only by a modest factor -- apparently the envelope
of the AGB star is expelled so easily by the companion, if the companion's 
mass is greater than 25\% of the AGB star's mass, that relatively little 
energy is extracted from the orbit.
\par However the two systems $\theta$ Tuc and V1379 Aql, which seem to
us to be much more probably genuine post-Algol systems, are themselves
going to suffer a common-envelope episode in the future, when star 2
evolves and initiates reverse RLOF. Here the mass ratio is more like
10:1, and we expect considerable spiral-in. The product could be either
a merger, with the white dwarf or hot subdwarf merging with the
hot-subdwarf core of the red giant, or a close WD + WD pair as in
$0957-666$ (Table 6). Since several such close pairs have been found
in the last few years (Marsh, Dhillon \& Duck 1995) we might favour the 
second alternative. But all these pairs contain WDs which are over 
$0.3\Msun$, which suggests substantially wider progenitors than the 
two post-Algols above. We therefore suspect that a merger is the more 
likely result for $\theta$ Tuc and V1379 Aql.
\par The merger of two WDs, or of a WD with the WD core of a red supergiant,
is sometimes seen as a possible explanation
of Type Ia SNe, but only if the combined mass exceeds the Chandrasekhar
mass. We are far short of that here. It seems more likely that the
result will simply be to ignite a helium-burning core, in a rather
extreme version of the helium flash. If not all of the hydrogen-rich
envelope of star 2 as a red giant is expelled during the common-envelope
phase, the result might even be as innocuous as a GK-clump giant, but with 
unusually rapid rotation. This would probably lose mass rapidly, and leave
an EHB star of $\ts 0.5\Msun$. Ironically, this is much the same as is
expected from a single red giant after the helium-flash, {\it provided}
all of the envelope is expelled at or near the helium flash.
\par The final system in Table 6 is AA Dor, an SDO star which was found
to be an eclipsing binary (Conti, Dearborn \& Massey 1981). This system 
has been interpreted as a post-common-envelope object, but it has the 
following problem, which relates to the fact that it appears to have a 
quite unusually low mass. Presumably the ancestor of star 1 was at least 
$1\Msun$ to start with, or it could hardly have evolved at all. It may
have filled its Roche lobe in an intial orbit of say $30\thin$d, by
which time its core would have had about the right mass. This would
no doubt have required very rapid mass transfer, in the manner of
Paczy{\'n}ski (1976), followed presumably by spiral-in. But it is very
hard to see how an envelope as massive as $0.75\Msun$ could have been
entirely ejected by a companion of only $0.05\Msun$, however rapidly
and far it spiralled in. The result would seem much more likely to be a 
complete merger.
\par Our tentative solution to this is Case AM evolution, as described
above. Even a fairly small companion, starting with say 
($1+0.05\Msun, 20\thin$d), should 
spin up a star to unusually rapid rotation as it climbs the giant branch. 
The mass loss before RLOF begins can be quite substantial, as already 
indicated for some of our computed systems. Mass loss with minimal angular 
momentum loss -- $\ra=R=I=0$ in equation (7) -- causes the orbit to expand: 
as a first approximation we can estimate $P\propto 1/M^2$, $M$ being the
combined mass. This is in practice an overestimate, since we do in fact 
expect some magnetic braking. But the expansion of the orbit
more-or-less at the same rate as the expansion of star 1 prolongs the
detached phase and the mass loss, so that the mass ratio may be only
say 6 ($0.3+0.05\Msun, \ts 60\thin$d) when RLOF begins. It will still be a
situation with probable dynamic RLOF, a common-envelope, and spiral-in.
But now the envelope is only $0.05\Msun$, and may be fairly readily
expelled by the low-mass companion without it spiralling in all the
way to a merger.
\par With our canonical model there was generally not enough mass
loss, and too much angular momentum loss, for this scenario, but
only by modest factors, $\tls 2$. While we have little doubt that
we could obtain good agreement by adjusting coefficients in the
formulae of Section 2, we conclude simply that $\ts 75\%$ mass loss prior 
to RLOF, and spiral-in afterwards, is a viable way of producing the
unusual system AA Dor, and more attractive than any alternative so
far. Although one could hardly call the system a post-Algol, it may
be influenced by much the same physical processes as pre-Algols,
Algols and post-Algols.
\section{Conclusions}
We have presented a simplistic formulation of mass loss driven by
dynamo activity, angular momentum loss driven by magnetic braking, 
and tidal friction, that in the first instance has no free parameters. 
It is calibrated to agree with the present-day Sun, and is scaled
according to the depth of convective envelope and the Rossby number in 
a realistic way. It appears to account for a fairly wide range of 
phenomena: the surprisingly slow rotation in BY Dra, the surprisingly 
low mass in the more evolved component of Z Her, and the surprisingly 
low angular momentum of Algols such as DN Ori and S Cnc. Variants of 
this model, in which some coefficients are altered by factors of up to 
$\ts 2$, may account for a wider range of observed objects, such as 
AS Eri, R CMa and AA Dor. Given that the dynamo activity which
is the basis of both the mass loss and the magnetic braking is an
inherently chaotic process, it is certainly not surprising that factors
of two or more should be necessary as between one system and another.
\par The same formulation can of course be applied to other classes of
object: contact binaries, cataclysmic variables, pre-cataclysmic systems, 
and low-mass X-ray binaries. We hope to pursue these in a future paper. 
Estimates have already been given by in Paper II.
\par An important improvement that will be necessary to understand
some further systems, such as cool double-subgiant binaries, is to
include star 2 within the formalism; for the present only star 1 is
allowed to be subject to these non-conservative processes. This will
be quite a major undertaking, since it will be necessary to solve for
both components {\it simultaneously}. However this is also necessary
if one is to follow the evolution of contact binaries, since an additional
model, for heat transport between the two components, is necessary there.
We hope to produce this in due course.
\par A different but very important way of pursuing the same topics is
to model these interactions in a fully 3-D stellar model, or pair of
models. For example, it would be desirable to model tidal friction in
such a way, to make a better estimate of the viscous timescale that
is incorporated in the constant factor of Equation (9). If MHD is
included, presumably just in the frozen-in approximation, then 3-D
calculations could also serve to calibrate the other non-conservative
processes. Most importantly, they could also give us insight into the
poorly-understood process of heat transport in contact binaries.
\par The DJEHUTY project is currently being developed at the Lawrence
Livermore National Laboratory, with a view to tackling the 3-D structure
of stars. At present grids of $\ts 10^8$ meshpoints are available, and
as computer power increases we hope to improve this to $\ts 10^{10}-
10^{11}$. Of course, one would not evolve such a 3-D model for several
Gigayears, but only for modest times like $\ts 1\thin$yr. This should
allow us to make a 1-D average of such processes as tidal friction, and
incorporate them in a 1-D code such as the one used here.

\acknowledgements

This work was undertaken as part of the DJEHUTY project at LLNL.  Work performed at LLNL 
is supported by the DOE under contract W7405-ENG-48.  

\end{document}